\documentclass[aps,prd,floatfix,twocolumn,superscriptaddress,longbibliography,showpacs,10pt]{revtex4-2}
\usepackage{amssymb, amsmath,color,mciteplus,graphicx,subfigure}
\usepackage{calc,epsfig,epstopdf,color,mciteplus,bm,mathrsfs}
\usepackage{times}
\usepackage{float}
\usepackage{lipsum}
\newcommand{\ket}[1]{| #1 \rangle} 

\usepackage[T1]{fontenc}
\usepackage{xspace}
\usepackage{ulem}

\begin{document}
	
\title{Variational thermal quantum simulation of the lattice Schwinger model}
\author{Xu-Dan Xie}
\affiliation{Guangdong Provincial Key Laboratory of Quantum Engineering and Quantum Materials, School of Physics and Telecommunication Engineering,
South China Normal University, Guangzhou 510006, China}

\author{Xingyu Guo}
\affiliation{Guangdong Provincial Key Laboratory of Nuclear Science, Institute of Quantum Matter, South China Normal University, Guangzhou 510006, China}
\affiliation{Guangdong-Hong Kong Joint Laboratory of Quantum Matter, Southern Nuclear Science Computing Center, South China Normal University, Guangzhou 510006, China}

\author{Hongxi Xing}
\affiliation{Guangdong Provincial Key Laboratory of Nuclear Science, Institute of Quantum Matter, South China Normal University, Guangzhou 510006, China}
\affiliation{Guangdong-Hong Kong Joint Laboratory of Quantum Matter,
Southern Nuclear Science Computing Center, South China Normal University, Guangzhou 510006, China}

\author{Zheng-Yuan Xue}
\affiliation{Guangdong Provincial Key Laboratory of Quantum Engineering and Quantum Materials, School of Physics and Telecommunication Engineering,
South China Normal University, Guangzhou 510006, China}
\affiliation{Guangdong-Hong Kong Joint Laboratory of Quantum Matter, Frontier Research Institute for Physics,
South China Normal University, Guangzhou 510006, China}

\author{Dan-Bo Zhang}
\email{dbzhang@m.scnu.edu.cn}
\affiliation{Guangdong-Hong Kong Joint Laboratory of Quantum Matter, Frontier Research Institute for Physics,
South China Normal University, Guangzhou 510006, China} \affiliation{Guangdong Provincial Key Laboratory of Quantum Engineering and Quantum Materials, School of Physics and Telecommunication Engineering,
South China Normal University, Guangzhou 510006, China}

\author{Shi-Liang Zhu}
\affiliation{Guangdong Provincial Key Laboratory of Quantum Engineering and Quantum Materials, School of Physics and Telecommunication Engineering,
South China Normal University, Guangzhou 510006, China}
\affiliation{Guangdong-Hong Kong Joint Laboratory of Quantum Matter, Frontier Research Institute for Physics,
South China Normal University, Guangzhou 510006, China}  
\collaboration{QuNu Collaboration}

\date{\today}
	
\begin{abstract}
Confinement of quarks due to the strong interaction and the deconfinement at high temperatures and high densities are a basic paradigm for understanding the nuclear matter. Their simulation, however, is very challenging for classical computers due to the sign problem of solving equilibrium states of finite-temperature quantum chromodynamical systems at finite density. In this paper, we propose a variational approach, using the lattice Schwinger model, to simulate the confinement or deconfinement by investigating the string tension. We adopt an ansatz that the string tension can be evaluated without referring to quantum protocols for measuring the entropy in the free energy. Results of numeral simulation show that the string tension decreases both along the increasing of the temperature and the chemical potential, which can be an analog of the phase diagram of QCD. Through numerical simulations on the classical computer, we demonstrate 
the potential of exploiting near-term quantum computers for investigating the phase diagram of finite-temperature and finite-density nuclear matters.
\end{abstract}
	
\maketitle

\section{Introduction}
The phase diagram of quantum chromodynamics~(QCD) can envelop our understanding on a variety of nuclear matter system under different conditions~\cite{sachdev1999quantum}. For instance, the quarks are confined into composite particles like hadrons due to the strong interaction~\cite{griffiths2020introduction}. At very high temperature deconfinement occurs and quarks get free, forming quark-gluon plasma~\cite{Shuryak:1980tp}. However, a large regime of the phase diagram is undetermined and its completion is one of the main challenges in physics. The challenge lies in the intrinsic difficulty for classical computers to solve the finite-temperature QCD system at finite density~\cite{peskin1995introduction}. The prevalent quantum Monte Carlo method can solve QCD problems at near-vanishing density, but it will meet the notorious sign problem at finite
density~\cite{mclerran1981quark,troyer2005computational}. Tensor networks can be free of the sign problem and achieve many meaningful results in the field of quantum many-body physics.   But to avoid a blowup of the computational cost (exponentially in $N$),
we need to truncate the matrices to a moderate bond dimension which introduces a truncation error ~\cite{orus2014practical}.

With rapid advances of quantum hardware, the promise of solving quantum many-body problems by quantum computing becomes increasingly within reach~\cite{liu2019variational,kokail2019self,zhang2021selected,dallaire2020application}. Notably, there are some progresses in exploiting quantum computing for solving nuclear physics that will be intrinsic hard with classical methods, such as real-time evolution~\cite{lloyd1996universal}, evaluation of parton distribution function~\cite{PhysRevResearch.2.013272,li2021partonic}, and so on~\cite{jordan2012quantum,klco2018quantum,martinez2016real,zohar2017digital,mueller2020deeply,czajka2021quantum,gallimore2022quantum}.  The Schwinger model is remarkable as a playground for simulating nuclear physics on the current quantum processors~\cite{kokail2019self,zhou2021thermalization,de2022quantum,halimeh2022tuning,yang2020observation}. As a model for the 1+1D quantum electrodynamics, the Schwinger model exhibits many interesting phenomena in common with 
QCD, such as confinement~\cite{buyens2016hamiltonian}, chiral symmetry breaking~\cite{hamer1982massive}, charge shielding~\cite{coleman1975charge}, and so on ~\cite{melnikov2000lattice,sachs2010finite,funcke2020topological}. In Ref.~\cite{buyens2016hamiltonian}, a tensor network approach has been adopted for studying the confinement properties of the Schwinger model at finite temperature and zero chemical potential, which can mimic the confinement of QCD but can be much simpler for quantum simulation.   

Investigation thermodynamics of the Schwinger model under varied temperatures and densities, as an analog to the phase diagram of QCD, can be feasible with thermal quantum simulation~(TQS)~\cite{zhang2021continuous,francis2021many,verdon2019quantum,wu2019variational,zhu2020generation}. The TQS aims to simulate finite-temperature quantum systems with quantum computers. The key of TQS is to prepare the Gibbs~(thermal) state describing the equilibrium state of a quantum system at a temperature. This can be achieved either by generating a purified thermal state~\cite{zhu2020generation,wu2019variational,chowdhury2020variational}, or preparing a density matrix as a mixing of pure states with a classical probability distribution~\cite{verdon2019quantum,liu2021solving,martyn2019product}. For the purpose of illuminating the feature of confinement or deconfinement, one can evaluate the string tension as a characteristic of the confinement strength. To calculate the string tension, however, relies on accessing to the free energy and thus the von Neumann entropy which is nontrivial to extract on a quantum computer~\cite{verdon2019quantum,martyn2019product}. In this regard, a choice of TQS method convenient for calculating the entropy is necessary.

In this paper, we adopt a variational quantum algorithm~(VQA) to investigate the lattice Schwinger model at varied temperatures and densities by evaluation of the string tension. The string tension is calculated as free energy difference with/without a pair of opposite charges at two ends of the chain. We use an ansatz convenient to calculate the free energy and thus the string tension, where the entropy can be analytically obtained. To illustrate the effectiveness of VQA, we perform numerical simulations on a classical computer. After testing the VQA at different temperatures and system sizes, we adopt the algorithm for exploring the dependence of string tension with both the temperature and the chemical potential. The results are consistent with theoretical predictions. Our work demonstrate the potential of exploiting near-term quantum computers for investigating the phase diagram of finite-temperature and finite density for nuclear matters. 

The rest of the paper is organized as follows. We first introduce the Schwinger model and the concept of string tension in Sec.~\ref{sec:model}. Then we propose a variational quantum algorithm for calculating the string tension in Sec.~\ref{sec:algorithm}. The numeral simulation results are presented in Sec.~\ref{sec:results}. Finally, conclusions are made in Sec.~\ref{sec:conclusion}.

\section{Finite-temperature Schwinger model and string tension} \label{sec:model}
In this section, we first present a lattice version of the Schwinger model. Then, we introduce the concept of string tension for the finite-temperature Schwinger model, as well as its expression by the free energy difference with and without a pair of trial charges at two ends of the chain. 
\subsection{The lattice Schwinger model}

The Schwinger model describes the 1+1D~(one spatial dimension +one time dimension) quantum electrodynamics with one flavor fermion. As a simply model of quantum gauge field theory, the Schwinger model provides a playground for studying a variety of physics, including the spontaneous
creation of electron–positron pairs from the vacuum~\cite{brezin1970pair}, the confinement of quark and antiquark~\cite{abdalla1991non,lowenstein1971quantum}, dynamical phase transition~\cite{byrnes2002density,buyens2017finite}, and so on. Remarkably, recent years it becomes a standard model for simulating both dynamical and static physical properties of lattice gauge theories on quantum computers.  

Let us start with a Hamiltonian description of the Schwinger model by fixing the temporal component of the vector potential as zero, $A_{0}(x)=0$. The Hamiltonian density is~\cite{zinn2002determination},  
\begin{eqnarray}
\label{Eq:Schwinger_model_continum}
H=&&\int dx  \Big[ \Psi ^{\dagger}(x)\gamma ^{0} \gamma ^{1}\big(-i\partial_{1}+g\hat{A}_{1}(x) \big)\Psi(x) \notag  \\ 
&&+m\Psi ^{\dagger}(x)\gamma ^{0}\Psi(x)+\frac{1}{2} \hat{E}^{2}(x)  \Big],  
\end{eqnarray}
where $\Psi(x)$ is a fermion field, $\hat{A}_{1}$ is the longitudinal vector potential and $m$ is the fermion mass. In 1+1D, the Dirac matrices read as $\gamma^{0}=\hat{\sigma}^{z}$ and $\gamma^{1}=i\hat{\sigma}^{y}$. The electric field satisfies $\hat{E}(x)=-\partial_{0}A_{1}(x)$ and the commutation relation $[\hat{A}_{1}(x), \hat{E}(x')]=-i\delta(x-x')$. By Gauss law there is a constraint between the electric field and the fermion density, 
$\partial_1E(x)=g\Psi(x)^\dagger\Psi(x)$. 

Following Ref.~\cite{kokail2019self}, a lattice formula of Eq.~\eqref{Eq:Schwinger_model_continum} can be obtained by the staggered lattice approach, which respectively recasts the two-conponent fermionic fields into the even and odd sites of the lattice, namely $\hat{\Phi}_{2j}=\sqrt{a}\hat{\Psi}_{e^{-}}(x_{2j})$ and $\hat{\Phi}_{2j-1}=\sqrt{a}\hat{\Psi}^{\dagger}_{e^{+}}(x_{2j-1})$, where $a$ is the lattice constant and $e^{\mp}$ represent fermion/antifermion. 
The gauge fields, now living on the links on the lattice, become $\theta _{j,j+1}=-agA(x_{j}+\frac{a}{2})$ and $L _{j,j+1}=\frac{1}{g}E(x_{j}+\frac{a}{2})$, which satisfy the commutation relation $[\theta _{j,j+1}, L _{j',j'+1}]=-i\delta_{j,j'}$. With the above recasting,  we get the Kogut-Susskind formulation for the Schwinger model of lattice size $N$~\cite{kogut1975hamiltonian}, 
\begin{eqnarray}
\label{ham_ks}
H=&&\frac{1}{2a} \sum_{j=1}^{N-1}[\hat{\Phi }_{j}^{\dagger }\hat{U}_{j,j+1}\hat{\Phi}_{j+1}+h.c.]
\notag \\
&&+m\sum_{j=1}^{N}(-1)^{j}\hat{\Phi }_{j}^{\dagger }\hat{\Phi}_{j}
+\frac{g^{2}a}{2} \sum_{j=1}^{N-1}\hat{L}_{j,j+1}^{2},
\end{eqnarray}
where $\hat{U}_{j,j+1}=e^{i\theta_{j,j+1}}$. The Gauss’law now becomes, 
\begin{eqnarray}
\hat{L}_{j,j+1}-\hat{L}_{j-1,j}=\hat{\Phi}_{j}^{\dagger}\hat{\Phi}-\frac{1-(-1)^{j}}{2}.
\end{eqnarray}
For an open boundary chain with a boundary condition $\hat{L}_{0,1}=\varepsilon$, the Gauss law reads,
\begin{eqnarray}
\label{eqL}
\hat{L}_{j,j+1}=\varepsilon+\sum_{l=1}^{j}\big [ \hat{\Phi }_{j}^{\dagger }\hat{\Phi}_{j}-\frac{1-(-1)^{l}}{2}\big ]. 
\end{eqnarray}
This means that the electric field can be determined by the fermion densities. 
The $\varepsilon$ term can be regarded as a background electric field. 
Moreover, with a gauge transformation of the fermion operators, $\hat{\Phi}_{j}\rightarrow {\textstyle \prod_{l=1}^{j-1}\hat{U}_{j,j+1}}\hat{\Phi}_{j}$, the $\hat{U}_{j,j+1}$ in Eq.(\ref{ham_ks}) can be absorbed and thus will not appear in the Hamilton. The gauge fields thus can be eliminated from the Hamiltonian at a price  that the Hamiltonian involves long-range density-density interaction. The Hamiltonian now can be expressed in terms of fermion degree of freedom~\cite{kokail2019self},   
\begin{eqnarray}
H=&&\varpi \sum_{j=1}^{N-1}(\hat{\Phi }_{j}^{\dagger }\hat{\Phi}_{j+1}+h.c.)
+m\sum_{j=1}^{N}(-1)^{j}\hat{\Phi }_{j}^{\dagger }\hat{\Phi}_{j}
\notag \\
&& +\frac{g^{2}a}{2} \sum_{j=1}^{N-1} \Big\{ \varepsilon+\sum_{l=1}^{j}\big [\hat{\Phi }_{j}^{\dagger }\hat{\Phi}_{j}-\frac{1-(-1)^{l}}{2}\big ] \Big\}^{2},    
\end{eqnarray}
where $\varpi=\frac{1}{2a}$. With a Jordan-Wigner transformation, $\hat{\Phi}_{j}={\textstyle\prod_{l=1}^{j-1}}(i\sigma_{l}^{z})\sigma_{j}^{-}$, where $\sigma^{\pm}=\frac{1}{2}(\sigma^{x}\pm i\sigma^{y})$, the model can be mapped to a qubit Hamiltonian, 
\begin{eqnarray}
\label{hspin}
H_\varepsilon=&&\varpi \sum_{j=1}^{N-1}[\hat{\sigma }_{j}^{+}\hat{\sigma}_{j+1}^{-}+h.c.  ]
+\frac{m}{2}\sum_{j=1}^{N}(-1)^{j}\hat{\sigma }_{j}^{z }
\notag \\
&& +\frac{g^{2}a}{2} \sum_{j=1}^{N-1}\Big\{\varepsilon+\frac{1}{2}\sum_{l=1}^{j}\big [ \hat{\sigma} _{l}^{z} +(-1)^{l}\big ]  \Big\}^{2},    
\end{eqnarray}
 
The qubit Hamiltonian in Eq.~\eqref{hspin} can be readily investigated on a quantum computer. It should be stressed that a complete elimination of the gauge field like the Schwinger model is not general for gauge theories. This makes the Schwinger model special as a starting model for studying lattice gauge theories by quantum computing, avoiding  the issue of dealing with gauge fields which can be resource demanding.  

Moreover, one can study the Schwinger model at finite chemical potential by adding an extra term $-\mu\sum_{j=1}^{N}\hat{\Phi }_{j}^{\dagger }\hat{\Phi}_{j}$ in the lattice Schwinger model, where $\mu$ is the chemistry potential~\cite{fetter2012quantum}. This turns the qubit Hamiltonian into a new Hamiltonian, 
\begin{equation}\label{eq:G_chemical}
G_\varepsilon(\mu)=H_\varepsilon-\frac{\mu}{2}\sum_{j=1}^{N}\hat{\sigma }_{j}^{z}.   
\end{equation}
By including the chemical potential term in $G_\varepsilon(\mu)$, one can investigate the finite-temperature Schwinger model at finite densities. 

\subsection{String tension}
One remarkable feature of the Schwinger model is confinement:
the potential energy between a fermion-antifermion pair grows linearly with the separation that no free fermion can be observed in this system~\cite{casher1974vacuum}. Such phenomena of confinement can mimic that of quantum chromodynamics, where a quark-antiquark pair are confined due to the strong interaction. The Schwinger model can be regarded as a toy model for investigating confinement related physics. 
A simple picture for understanding the confinement is to look at the large $m$ limit. If we put one fermion and one antifermion on the bare vacuum with a separation $r$, then there will be an electric field of strength $g$ between them due to the Gauss law. The potential energy due to the presence of electric field grows linearly with the separation, and the force between two fermions thus is a constant. In general, one may imagine there is a string connecting two confined fermions, and the strength of the confinement can be characterized by the string tension. At finite temperatures, thermal fluctuations of fermion-antifermion pairs
will reduce the strength of the string tension and the confinement will be weakened. For the Schwinger model, it can be analyzed that the string tension will diminish exponentially with the temperature~\cite{Fischler_1979,buyens2016hamiltonian}. This is different from the QCD, where there is a confinement-deconfinement transition at a finite temperature~\cite{Pisarski_1982,fodor2004critical,d2003finite,allton2003equation}. 

The string tension is a static property and its temperature-dependence can be calculated by quantum statistical mechanics~\cite{hamer1982massive}. Let us consider a chain described by the Schwinger model at a temperature $T$. Firstly, put a pair of
 fermion and antifermion with trial charge $\pm\varepsilon g$ at the left end of the chain. Then, move the antifermion to the other end of the chain. The free energy will increase as work has been done in this process. We can define the string tension as a force that contributes to the work. For an original system described by a Hamiltonian in Eq.~\eqref{hspin} with $\varepsilon=0$, the final Hamiltonian turns to be $H_\varepsilon$, where the setup of a pair of fermion and antifermion at two ends can correspond to a background electric field $\varepsilon g$. Thus, it suffices to calculate the difference of free energies for $H_0$ and $H_\varepsilon$. 
 
 In addition, we should subtract a contribution solely from the trial charges. This can be achieved by dropping terms only related to $\varepsilon$ in Eq.~\eqref{hspin}, namely $f_\varepsilon=\frac{g^2a(N-1)}{2}(\varepsilon^2-\frac{\varepsilon}{2})$. 
 The string tension can be written as~\cite{buyens2016hamiltonian}, 
\begin{eqnarray}
\label{tension}
\sigma_{\varepsilon}(\beta)=\frac{1}{Nga}\big(F_{\varepsilon}(\beta)-F_{0}(\beta)-f_\varepsilon\big),
\end{eqnarray}
where $\beta=\frac{1}{T}$ is the inverse temperature and $F_{\varepsilon}(\beta)$ is the free energy for the system of Hamiltonian $H_\varepsilon$. The same procedure can also apply for the Schwinger model with finite chemical potential by replacing $H_\varepsilon$ with $G_\varepsilon(\mu)$.

\section{Variational quantum algorithm for evaluating the string tension} \label{sec:algorithm}
We now investigate how the string tension of the finite-temperature Schwinger model can be evaluated on a quantum computer. We first discuss the special role of entropy for quantum simulating finite-temperature quantum systems and then propose a variational quantum algorithm for evaluating the string tension. 

\subsection{Motivation}
As seen in Eq.~\eqref{tension}, the key of evaluating the string tension, as a difference of free energies, lies in the capacity of evaluating the free energy efficiently. In equilibrium thermodynamics, the system at an inverse temperature $\beta$ can be described by a Gibbs~(thermal) state~\cite{bogolubov2009introduction}, 
\begin{equation}
    \rho(\beta)=\frac{1}{Z(\beta)}e^{-\beta H}, ~~ Z(\beta)=\text{Tr}(e^{-\beta H}). 
\end{equation}
Let us write $H\ket{\varphi_n}=E_n\ket{\varphi_n}$ with $E_n$ the $n$-th eigenvalue for the corresponding eigenstate $\ket{\varphi_n}$. Then the Gibbs state can be regarded as a mixed state that the finite-temperature has a classical probability of $e^{-\beta E_n}/Z(\beta)$ in the pure state $\ket{\varphi_n}$.  

The free energy of the Gibbs state can be expressed as~\cite{bogolubov2009introduction}, 
\begin{eqnarray} \label{eq:F_E_S}
F(\beta)=E(\beta)-TS(\beta),
\end{eqnarray}
where $E(\beta)=\text{Tr}[\rho(\beta)H]$ is the average energy and  $S(\beta)=-\text{Tr}[\rho(\beta)\log\rho(\beta)]$ is the von Neumann entropy. Moreover, the Gibbs state $\rho(\beta)$ minimizes the free energy. In this regard, the expression of free energy in Eq.~\eqref{eq:F_E_S} sets a variational principle for preparing the Gibbs state by minimizing the free energy. 

The quantum statistical physics involving preparing the Gibbs state as well as investigating the thermodynamical properties in general is hard for classical computers due to the exponential increasing of Hilbert space~\cite{kempe2006complexity}. For the phase diagram of QCD, for instance, it is known that the quantum Monte Carlo method can apply well for the regime of near-zero chemical potential at different temperatures~\cite{de2002qcd}, but its accuracy for exploration of a regime of non-zero chemical potential cannot be guaranteed. By including a nonzero chemical potential term as in Eq.~\eqref{eq:G_chemical}, a nonzero baryon density will appear and there is an imbalance between quark and antiquark. Consequently, non-positive-definiteness of the weight function will arise in transforming a quantum problem into a form resembling a classical statistical mechanics problem, which leads to the sign problem in quantum Monte Carlo method~\cite{splittorff2007qcd}. It is a NP hard problem to solve the sign problem for simulating quantum many-body systems on a classical computer~\cite{troyer2005computational}. This intrinsic difficulty, however, can be avoided if we use a quantum computer to simulate quantum many-body systems. 

Nevertheless, even with a quantum computer, there is a subtlety for simulating finite-temperature quantum systems involving the entropy~\cite{lindblad1991quantum}. As expressed in Eq.~\eqref{eq:F_E_S}, evaluating the free energy requires first to prepare a Gibbs state and then measure the average energy and the entropy. Once the Gibbs state is available on a quantum computer, the average energy can be obtained efficiently by decomposing the Hamiltonian as a sum of Pauli operators and performing each Pauli measurement alone. As shown in Eq.~\eqref{H_pauli}, the number of Pauli operators for the Schwinger model is proportional to $O(N^3)$~(see Appendix A for details), which indicates that the average energy can be measured efficiently.

However, estimation of the entropy, which cannot be taken as a quantum average of an Hermitian operator, meets a difficulty: in general it requires to diagonalize the full density matrix $\rho(\beta)$. While there are some quantum protocols to calculate the entropy as a function of density matrix $\rho$, either with a collective measurement on multiple copies of $\rho$~\cite{islam2015measuring} or with the random measurements~\cite{klich2006measuring}, those protocols access to the trace of a power $\rho$ known as the Rényi entropy~\cite{brydges2019probing}. It is still unknown how to compute the von Neumann entropy efficiently, although there are some estimations of the von Neumann entropy with proper approximations~\cite{audenaert2007sharp,gilyen2019distributional,acharya2019measuring}. In the next subsection, we will adopt a variational approach for preparing the Gibbs state, with an ansatz that the entropy can be readily calculated on a classical computer.    

\subsection{Variational quantum algorithm}
The variational principle for preparing a Gibbs state is to minimize the free energy with regard to the parameterized mixed state~\cite{verdon2019quantum}. Let us denote the parameterized mixed state as $\rho(\boldsymbol{\omega})$ with a parameter set $\boldsymbol{\omega}$. For a given inverse temperature $\beta$, the variational free energy then becomes a function of $\boldsymbol{\omega}$, 
\begin{equation}\label{eq:F_variational}
    F(\boldsymbol{\omega},\beta)=\text{Tr}[\rho(\boldsymbol{\omega})H]-\beta^{-1}S(\boldsymbol{\omega}), 
\end{equation}
where the entropy is $S(\boldsymbol{\omega})=-Tr[\rho(\boldsymbol{\omega})\log\rho(\boldsymbol{\omega})]$. For the purpose of calculating entropy, we adopt an ansatz to parameterize the target Gibbs state with a parameter set $\boldsymbol{\omega}=(\theta,\phi)$, 
\begin{equation}\label{eq:ansatz}
    \rho(\boldsymbol{\omega})=U(\phi)\rho_0(\theta)U^\dagger(\phi),
\end{equation}
where $\rho_0(\theta)$ is an initial mixed state easy to prepare on the quantum computer, and $U(\phi)$ is an unitary implemented with a parameterized quantum circuit. 

\begin{figure}[tbp]
	\centering
	\includegraphics[width=0.95\linewidth]{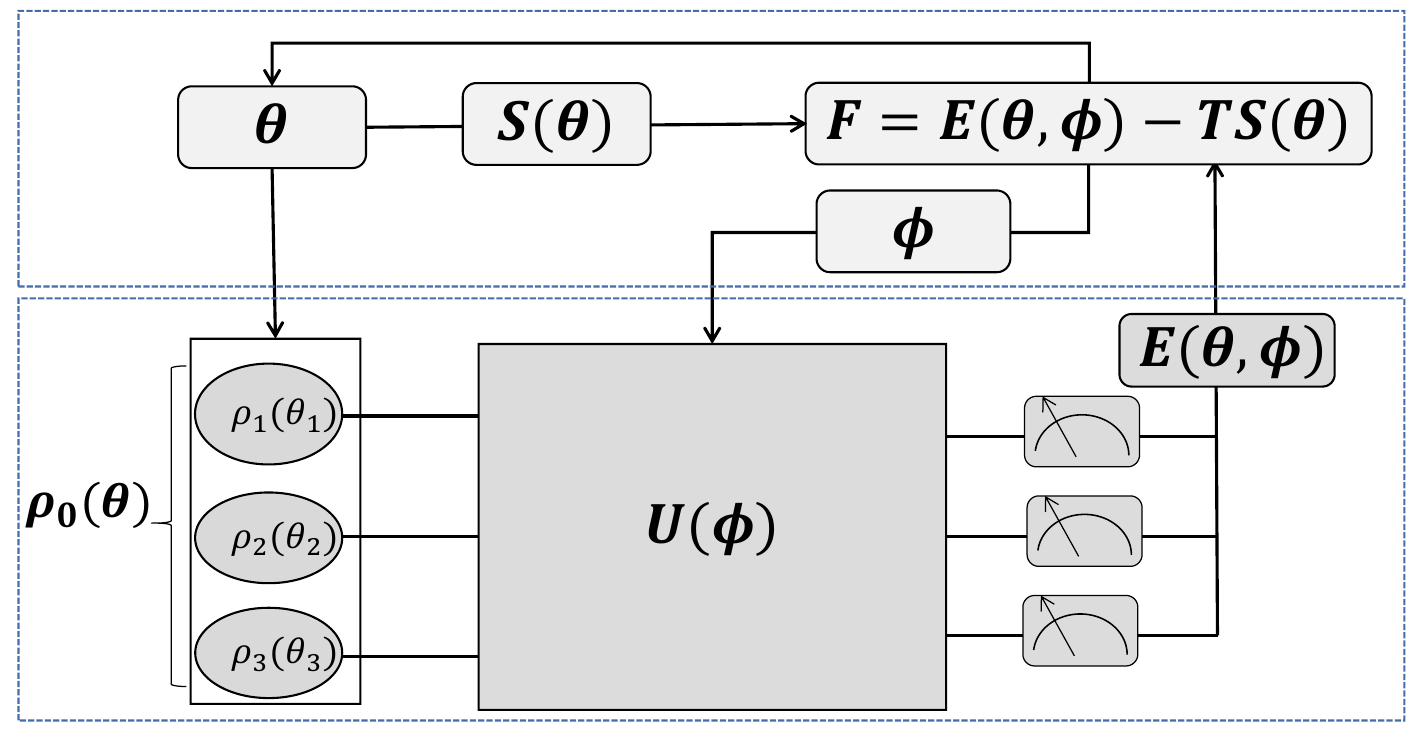}
	\caption{ A pictorial representation of the variational quantum algorithm, which can prepare thermal states and compute    the corresponding free energy. In quantum device, we design an initial product state $\rho_{0}(\theta)$  and the 
    unitary quantum neural network $U(\phi)$ parameterized by a serious
    of variable parameters. Using classical computers, the variable parameters $(\theta,\phi)$ are updated and optimized in order to minimize the free energy. }\label{Fig:illustration}
\end{figure}

As $\rho_0(\theta)$ and $\rho(\boldsymbol{\omega})$ are related by an unitary transformation, the eigenvalues will not change and so as to the entropy. Thus, the entropy depends only on the parameters $\theta$. A fully parameterization of the eigenvalues of the Gibbs state into the initial mixed state $\rho_{0}(\theta)$, however, requires an exponential number of parameters. Nevertheless, the goal is to prepare the Gibbs state rather than its each component, and it is not necessary to access each eigenstate $\ket{\varphi_n}$ with a weighting $e^{-\beta E_n}/Z(\beta)$ accurately. Therefore, one can parameterize the initial mixed state with much fewer parameters, e.g., a polynomial of the number of qubits. As shown in the FIG.\ref{Fig:illustration}, We adopt $\rho_0(\theta)$ as the initial state, which is composed of $N$ subsystems independent with each other. As the state space of a composite physical system is the tensor product of the state spaces of the component physical systems~\cite{nielsen2002quantum}, the initial state can be written as in a form of tensor product

\begin{eqnarray}
\label{eq:rho_0_ansatz}
\mathcal\rho_{0}(\theta)=\otimes_{i=1}^{N}\rho _{i}(\theta_{i}),
\end{eqnarray}

In Ref.~\cite{martyn2019product}, this ansatz, together with the parameterized unitary in Eq.~\eqref{eq:ansatz}, is called as product spectrum ansatz~(PSA) since it approximates the spectrum of the Gibbs state with a product structure. The PSA may be justified by the locality of temperature~\cite{kliesch2014locality}: correlations between two regimes of a quantum system decreases significantly with the separation, and the decreasing is quicker for higher temperatures. The PSA in Eq.~\eqref{eq:ansatz} starts in $\rho_{0}$ with a completely localized description of temperature and then encode the correlation by an unitary operator $U(\phi)$.  In our numeral simulation, we find that the PSA suffices for simulating the temperature-dependence string tension of the finite-temperature Schwinger model. 

With an initial state as a product state, the entropy for both the initial state and the final state $\rho(\boldsymbol{\omega})$ can be calculated as, 
\begin{eqnarray}
\label{Eq4} 
\mathcal S(\rho(\boldsymbol{\omega}))=S(\rho_0(\theta))=\sum_{i}^{N}S_{i}(\rho_{i}(\theta_{i})), 
\end{eqnarray}
where each entropy $S_{i}(\rho_{i}(\theta_{i}))$ can be obtained analytically. 

Since variational free energy can be readily evaluated, we can now minimize the variational free energy in Eq.~\eqref{eq:F_E_S} by finding the optimized parameter set $\boldsymbol{\omega}$ with a hybrid quantum-classical optimization. An illustration of the variational quantum algorithm can be found in Fig.~\ref{Fig:illustration}. In the below, we outline some details of the ansatz. 

\begin{figure}[tbp]
	\centering
	\includegraphics[width=0.8\linewidth]{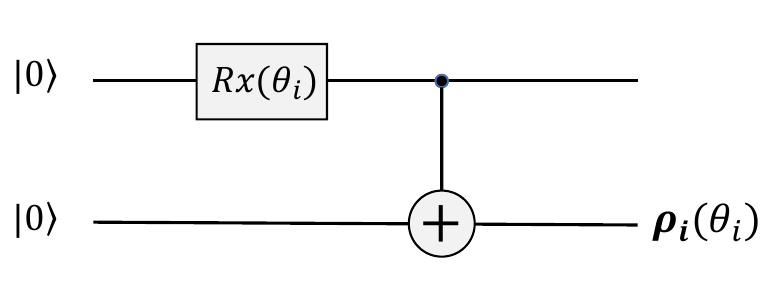}
	\caption{The circuit of preparing one-qubit mixed state as a subsystem. $Rx(\theta_{i})=e^{i\theta_i\sigma^x_i}$ is a one-qubit rotation.}\label{Fig:one_qubit_mix}
\end{figure}

\textbf{Preparing the initial state.} The initial state $\rho(\theta)$ is a product of single-qubit mixed state and it suffices to illustrate how to prepare a single-qubit mixed state. This can be achieved by preparing a two-qubit entangled pure state and trace one of the two qubits. For this, we adopt an ancilla qubit for a system qubit.  With a circuit illustrated in Fig.~\ref{Fig:one_qubit_mix}, the two-qubit entangled state becomes, 
$\cos\theta_{i}\left | 00 \right \rangle+\sin \theta _{i}\left | 11 \right \rangle.$ By tracing out the first qubit~(by ignoring the first qubit on the quantum computer), we get a mixed state(see Appendix B for details)
\begin{eqnarray}
\rho_i(\theta _{i})=\sin^{2}\theta _{i}|0\rangle\langle 0|+\cos^{2}\theta _{i}|1\rangle\langle 1|.
\end{eqnarray}
The entropy can be obtained analytically as, 
\begin{eqnarray}
    \label{Eq:entropy_one_qubit}
    S_{i}(\rho _{i}(\theta_i))=-\sin^{2}\theta_{i}\log(\sin^{2}\theta_{i})-\cos^{2}\theta_{i} \log(\cos^{2}\theta_{i}). \nonumber\\
\end{eqnarray}

For each system $i$-th qubit, we can attach an ancilla qubit for preparing the initial state $\rho_i(\theta_i)$. For a system of $N$ qubits these should be $N$ ancilla qubits. Thus, it requires $2N$ qubits to simulate a system of $N$ qubits. 

\begin{figure}[tbp]
\centering
\includegraphics[width=0.95\linewidth]{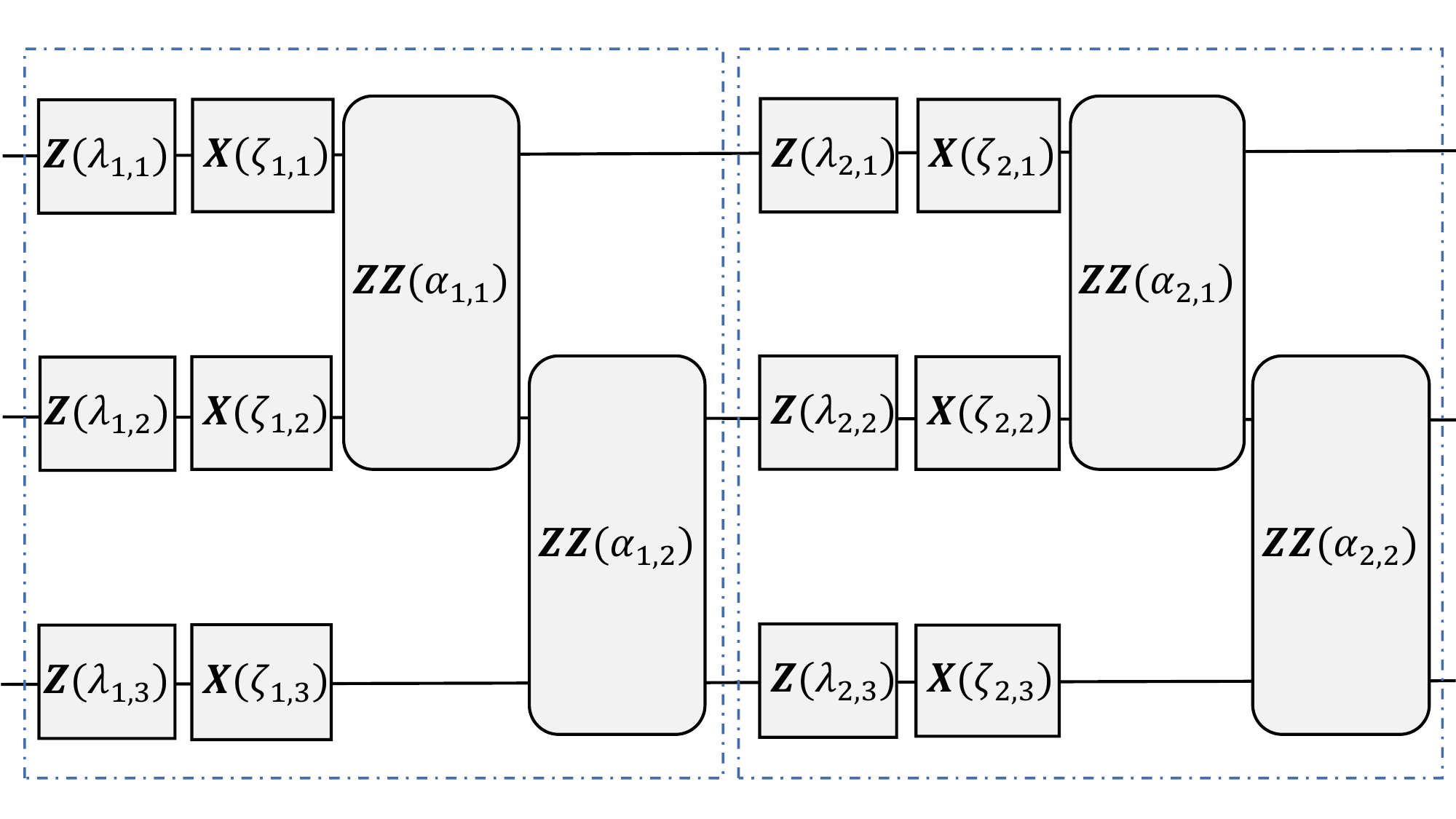}
\caption{ A depiction of the parameterized unitary for a system of
three qubits in the case $p=2$. }\label{Fig: unitary_circuit}
\end{figure}

\textbf{Construct the unitary operator $U(\phi)$}. The unitary operator is parameterized with only local unitary transformations with one-qubit and two-qubit rotations~\cite{nielsen2002quantum}, as illustrated in Fig.~\eqref{Fig: unitary_circuit}. Concisely, the circuit involves $p$ blocks, each block includes two layers of single-qubit rotations and one layer of two-qubit rotations, namely(with a parameter set $\phi=(\lambda,\zeta,\alpha)$), 
\begin{eqnarray}
\label{Eq:unitary} 
U(\phi)=\prod_{l=1}^{p}e^{-iH_{zz}(\alpha _{l})}e^{-iH_x({\lambda _{l}})}e^{-iH_{z}(\zeta_{l})},\nonumber\\
\end{eqnarray}
where $H_{z}(\lambda _{l})=\sum_{i}^{N}\lambda_{l,i}Z_{i}$,  $H_{x}(\zeta _{l})=\sum_{i}^{N}\zeta_{l,i}X_{i}$, $H_{zz}({\alpha }_{l})=\sum_{i}^{N-1} {\alpha}_{l,i}Z_{i}Z_{i+1}$,
$\lambda_{l,i}$, $\zeta_{l,i}$ and $\alpha_{l,i}$ are parameters at the $l$-th block.  We define $p$ as the circuit depth. In general, the larger $p$, the better the expressivity  of the unitary operator for representing the target state.

\section{Simulation results}\label{sec:results}

\begin{figure}[htbp]
	\centering
	\includegraphics[width=1\linewidth]{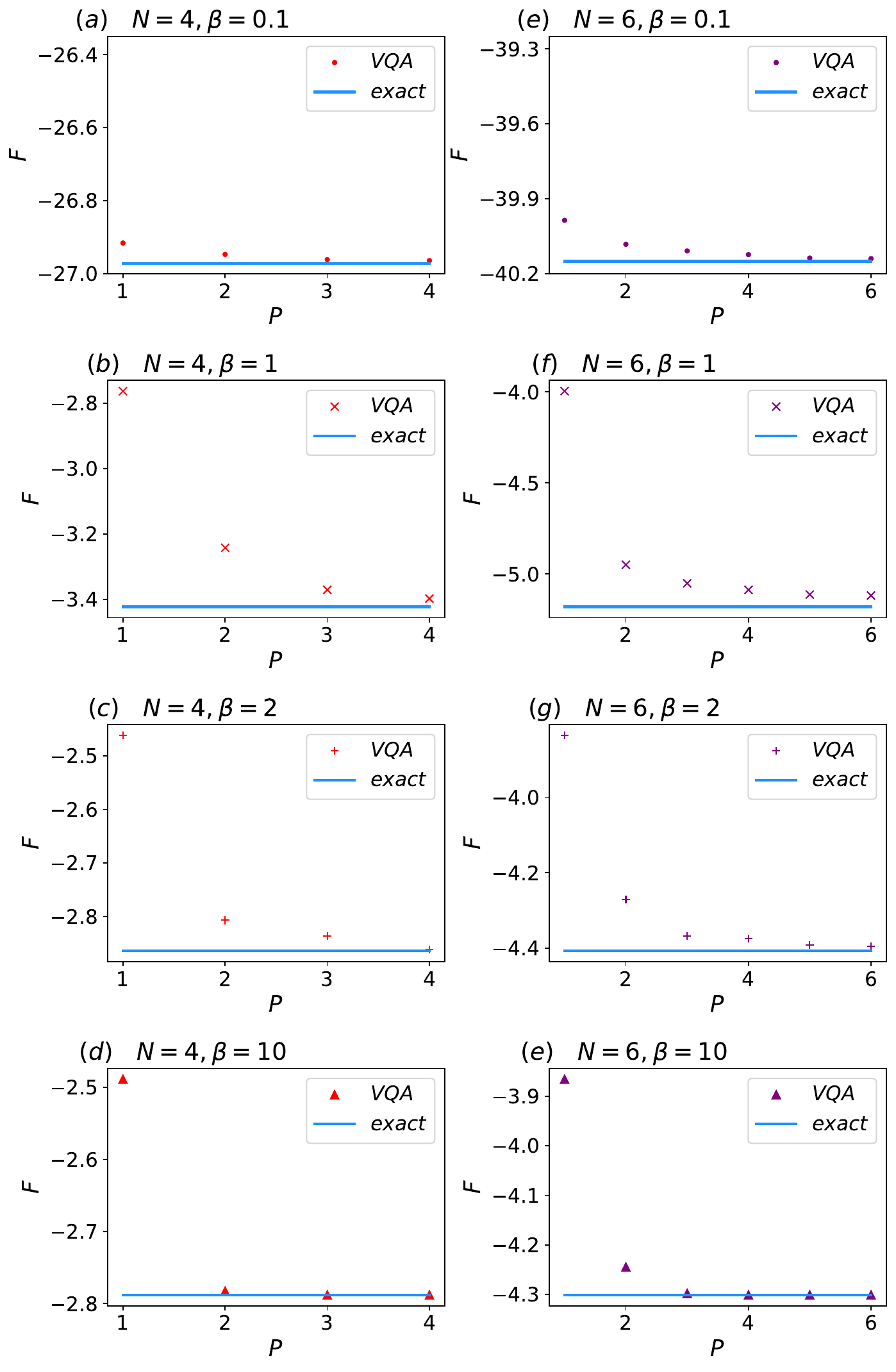}
	\caption{The free energy of the Schwinger model in the case $m=1,g=1, \varpi=1$. Left: the system sizes N=4. Right: the system sizes N=6. As the depth $p$ of the   increases, the free energies calculated by the quantum variational algorithm will converge close to the exact values.}\label{FigF_P}
\end{figure}
\begin{figure*}[htbp]
	\includegraphics[width=0.85\textwidth]{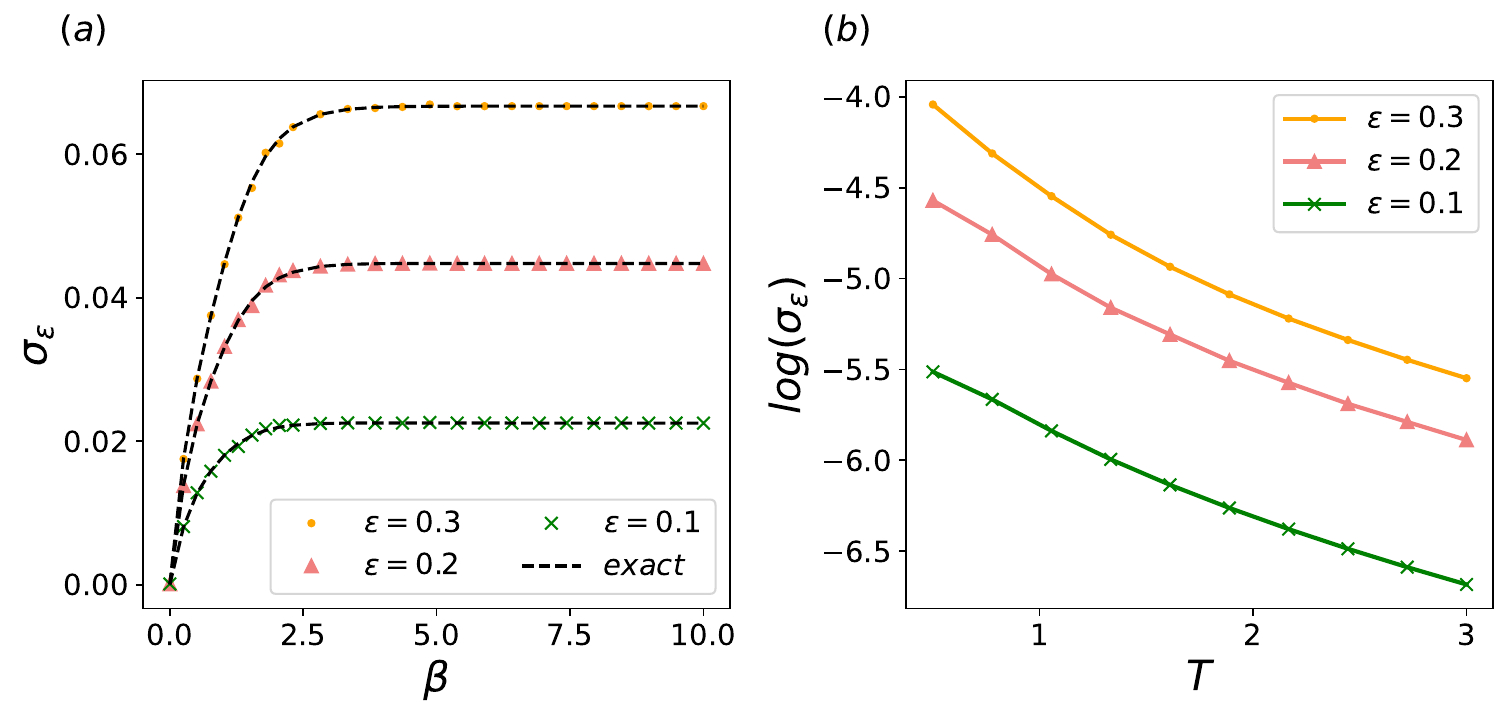}
	\caption{The string tension for different $\varepsilon$ in the case $m=1,g=1,\varpi=1$,$N$=6. (a) the string tension as a function of inverse temperature $\beta$. (b) logarithm of the string tension as a function of temperature $T$. }\label{ten6_T}
\end{figure*}

\begin{figure*}[htbp]\centering
	\includegraphics[width=0.85\textwidth]{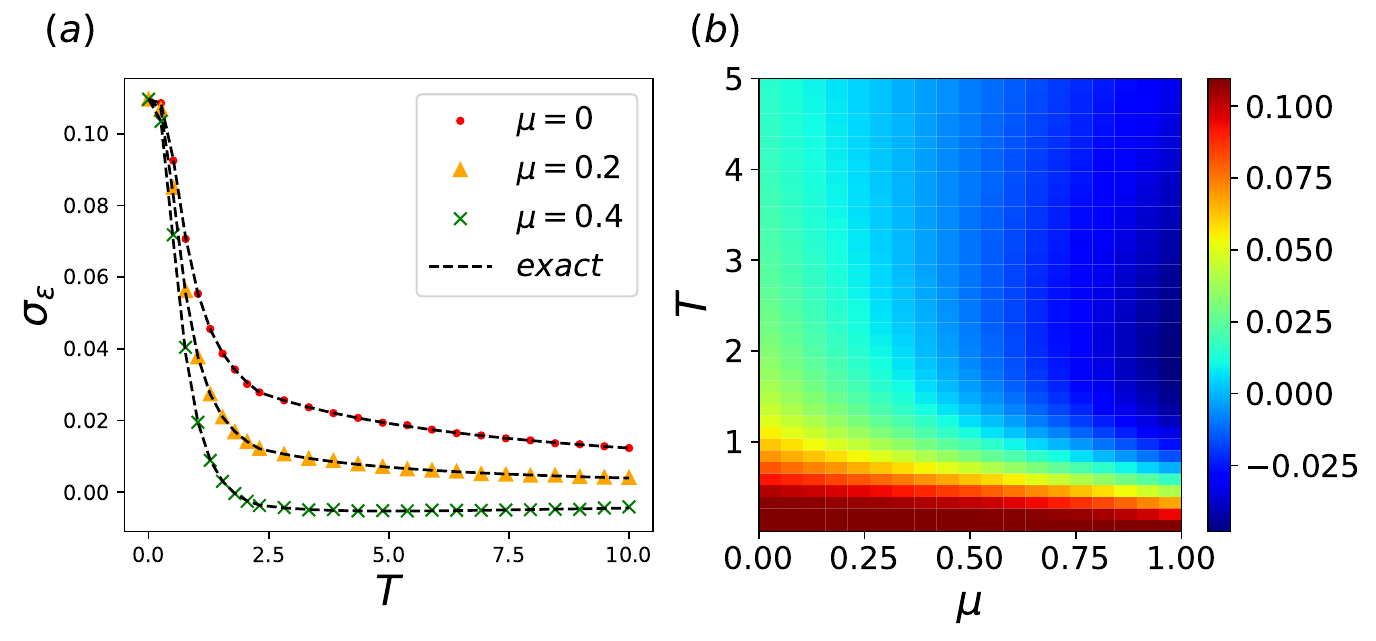}
	\caption{The string tension in the case $\varepsilon=0.5, m=1, g=1,\varpi=1,N=6$. (a) At different $\mu$, the string tension as a function of the temperature $T$; (b) The string tension as a function of the temperature $T$ and the chemical potential $\mu$. }\label{ten_T_u}
\end{figure*}
In order to illustrate the effectiveness of VQA, we perform numerical simulations on a classical computer. We now present simulation results for the finite-temperature Schwinger model. Specially, to evaluate the string tension at a given temperature, one can apply the
variational quantum algorithm to calculate the free energy of the Schwinger model in Eq.~\eqref{hspin} or Eq.~\eqref{eq:G_chemical} both for zero and nonzero $\varepsilon=0$. Then the string tension is obtained by the Eq.~\eqref{tension}. The numeral simulation is performed with the open source package \textit{QuTip}~\cite{johansson2012qutip}.

We first test the variational quantum algorithm for the finite-temperature Schwinger model at different system sizes and with different circuit depth $p$. The optimal variational free energies are compared with the exact values, as shown in Fig.~\ref{FigF_P}. As the depth $p$ of the parameterized quantum circuit increases, the optimal free energies calculated by the quantum variational algorithm will converge close to the exact values.  This can be expected as the larger $p$, the more expressive for the parameterized quantum circuit~\cite{ho2019efficient}. By comparing the results at different temperatures, we find that VQA can achieve more accurate results with small $p$ at high temperature~($\beta=0.1$) and at low temperature~($\beta=10$), while at intermediate temperatures it requires larger $p$ to make the results converge. The relatively easier for preparing thermal states at the high-temperature and the low-temperature limits can be expected~\cite{martyn2019product}. For the low-temperature limit $T<\Delta$~($\Delta$ is the gap of the system), the distribution concentrates to the ground state, with a small distribution on a few of low-lying excited states. With a very small entropy in the initial mixed state $\rho_0(\theta)$, the circuit depth $p$ in $U(\phi)$ is mainly required to  express the ground state faithfully rather than a set of eigenstates. At the high-temperature limit, the correlation length becomes very short, and the thermal state can be well described by the ansatz in Eq.~\eqref{eq:ansatz} with small $p$ and large entropy in the initial mixed state $\rho_0(\theta)$. 

We then use the quantum variational algorithm for investigating the thermodynamical properties of the Schwinger model by investigating its temperature-dependence of the string tension. As shown in Fig~\ref{ten6_T}a, the string tension decreases quickly at small $\beta$, meaning that the string tension is much weakened due to thermal fluctuation at large temperatures. The string tension will vanish at $T\rightarrow \infty$. The quick decreasing of string tension with increasing temperatures holds for different $\varepsilon$. To further reveal the behavior of decreasing string tension, we redraw the figure in Fig~\ref{ten6_T}b for a regime of temperatures that the string tension starts to decrease. It is shown that  the decreasing is almost exponentially, which is more obvious for smaller $\varepsilon$. The exponentially decreasing behavior is consistent with that of theoretical results~\cite{buyens2016hamiltonian}. 

Finally, we adopt the VQA for calculating the string tension of the Schwinger model by tuning both the temperature and the chemical potential. In Fig.~\ref{ten_T_u}a, the string tensions are investigated for different chemical potentials. It can be seen that the string tension is smaller for larger chemical potential at the same temperature. The dependence of the string tension with both the temperature and the chemical potential is then revealed in the Fig.~\ref{ten_T_u}b. The decreasing of string tension both along the direction of raising the temperature and the increasing of chemical potential in a large regime can be an analog to that of the QCD~\cite{Pisarski_1982,fodor2004critical,d2003finite,allton2003equation}. However, as the string tension only vanishes at $T\rightarrow\infty$, there is no confinement-deconfinement transition as in the QCD phase diagram. Nevertheless, the Schwinger model may be regarded as a model that can simulate how physics of confinement are affected by the temperature and the finite-density. In addition, the string tension can become negative when both the chemical potential is large and the temperature is moderate high. In such a regime, there can be a surplus of fermions with repulsive interactions, which are screened due to thermal fluctuation of fermion-antifermion pairs.

\section{Discussion and Conclusions}\label{sec:conclusion}

We have applied the variational quantum algorithm for simulating the Schwinger model at finite temperature and finite density. While the VQA approach can be of general purpose, an extension of our method for investigating the QCD phase diagram still requires further developments of some important ingredients. On one hand, it calls for efficient schemes for encoding non-Abelian gauge fields into qubits. While gauge field in 1+1D non-Abelian Schwinger model can still be eliminated~\cite{Atas:2021ext}, at higher dimensions the elimination (either Abelian or non-Abelian ) is impossible, although the number of qubits may be reduced by exploiting the Gauss law~\cite{banuls_17,klco_20}. On the other hand, the scalability of VQA for large-size quantum systems should be guaranteed as one expects that the number of qubits required for studying QCD problems shall be comparatively large. One obstacle is the problem of vanishing gradients, known as barren plateaus~\cite{mcclean_barren_2018,bittel2021training}, that prevents efficient optimization for large-size quantum circuits. Among some solutions to avoid the issue of barren plateaus~\cite{haug2021optimal,grant2019initialization,sack2022avoiding,kulshrestha2022beinit,anand2021natural}, One notable solution is to design proper ansatz specific for lattice models involving gauge fields. 

In addition, it should be pointed out that we have used a pair of trial charges at two ends of the chain, which is a bare fermion pair state, to probe the string tension. A possible improvement may refer to the pioneering research of K.G. Wilson~\cite{Wilson_1974}, which uses expectation value of the Wilson line to evaluate of the potential~(string tension) between the heavy quark-anti-quark pair. It is expected such an approach may be adopted on a quantum computer by mapping all to qubits and evaluating the Wilson line with regard to the Gibbs state.
 
In summary, we have applied a VQA to simulate the lattice Schwinger model at varied temperatures and densities. The physics of confinement and deconfinement has been investigated by evaluating the string tension, which is convenient to calculate by using a product-spectrum ansatz. By numerical simulations on classical computers, we have tested the VQA at varied system sizes, temperatures and densities, which fit well with exact diagonalization. Finally, we obtained by classical numeral simulation the dependence of string tension with temperatures and densities in a large regime. Our work 
shows the potential for simulating the phase diagram of nuclear matters on near-term quantum processors. 

\begin{acknowledgements}
This work was supported by the Guangdong Major Project of Basic and Applied Basic Research No. 2020B0301030008, the National Natural Science Foundation of China (Grant No.12005065, No. 12022512, No. 12035007), the Guangdong Basic and Applied Basic Research Fund (Grant No.2021A1515010317), and the Guangdong Provincial Key Laboratory (grant no. 2020B1212060066). 
\end{acknowledgements}

\appendix
\section{} \label{Ham_decompose}
From Eq(\ref{hspin}), the Schwinger model Hamiltonian can be written as
\begin{eqnarray}
H_\varepsilon=H_1+H_2+H_3  
\end{eqnarray}
where
\begin{eqnarray}
H_1&&=\varpi \sum_{j=1}^{N-1}[\hat{\sigma }_{j}^{+}\hat{\sigma}_{j+1}^{-}+h.c.]
=\varpi \sum_{j=1}^{N-1}[\hat{\sigma }_{j}^{+}\hat{\sigma}_{j+1}^{-}+\hat{\sigma }_{j+1}^{+}\hat{\sigma}_{j}^{-}] \notag \\
&&=\frac{\varpi}{2} \sum_{j=1}^{N-1}[\hat{\sigma }_{j}^{x}\hat{\sigma}_{j+1}^{x}+\hat{\sigma }_{j}^{y}\hat{\sigma}_{j+1}^{y}] 
\end{eqnarray}
\begin{eqnarray}
H_2&&=\frac{m}{2}\sum_{j=1}^{N}(-1)^{j}\hat{\sigma }_{j}^{z } 
\end{eqnarray}
\begin{eqnarray}
H_3&&=\frac{g^{2}a}{2} \sum_{j=1}^{N-1}\Big\{\varepsilon+\frac{1}{2}\sum_{l=1}^{j}\big [ \hat{\sigma} _{l}^{z} +(-1)^{l}\big ]  \Big\}^{2} \notag \\
&&=\frac{g^{2}a}{2}\sum_{j=1}^{N-1}\Big\{\varepsilon^{2}+\varepsilon\sum_{l=1}^{j}\big [ \hat{\sigma} _{l}^{z} +(-1)^{l}]  \notag \\
&&+\frac{1}{4}\sum_{l=1}^{j}\big [ \hat{\sigma} _{l}^{z}+(-1)^{l}]\sum_{l=1}^{k}\big [ \hat{\sigma} _{k}^{z}+(-1)^{k}]\Big\} \notag \\
&&=\frac{g^{2}a}{2}(N-1)\varepsilon^{2}-\frac{g^{2}a}{4}(N-1)\varepsilon
+ \frac{g^{2}a}{2}\varepsilon\sum_{j=1}^{N-1} \sum_{l=1}^{j} \hat{\sigma} _{l}^{z} \notag \\
&&+ \frac{g^{2}a}{4}\sum_{j=1}^{N-1} \sum_{l=1}^{j} \sum_{l=1}^{k}(-1)^{k}\hat{\sigma} _{l}^{z} + \frac{g^{2}a}{8}\sum_{j=1}^{N-1} \sum_{l=1}^{j} \sum_{k=1}^{j}\hat{\sigma} _{l}^{z}\hat{\sigma} _{k}^{z} \notag \\
&&+\frac{g^{2}a}{8}\sum_{j=1}^{N-1} \sum_{l=1}^{j} \sum_{k=1}^{j}(-1)^{(k+l)} \notag \\
&&=\frac{g^{2}a}{2}(N-1)(\varepsilon^{2}-\frac{\varepsilon}{2}) + \frac{g^{2}a}{2}\varepsilon\sum_{j=1}^{N-1} \sum_{l=1}^{j} \hat{\sigma} _{l}^{z} \notag \\
&&+ \frac{g^{2}a}{4}\sum_{j=1}^{N-1} \sum_{l=1}^{j} \sum_{k=1}^{j}(-1)^{k}\hat{\sigma} _{l}^{z} 
+  \frac{g^{2}a}{8}\sum_{j=1}^{N-1} \sum_{l=1}^{j} \sum_{k=1}^{j}\hat{\sigma} _{l}^{z}\hat{\sigma} _{k}^{z}
\notag \\
\end{eqnarray}
So the Schwinger model Hamiltonian can be can be written as
\begin{eqnarray}
\label{H_pauli}
&&H_\varepsilon=\frac{\varpi}{2} \sum_{j=1}^{N-1}[\hat{\sigma }_{j}^{x}\hat{\sigma}_{j+1}^{x}+\hat{\sigma }_{j}^{y}\hat{\sigma}_{j+1}^{y}]+ \frac{m}{2}\sum_{j=1}^{N}(-1)^{j}\hat{\sigma }_{j}^{z }\notag \\
&&+\frac{g^{2}a}{2}(N-1)(\varepsilon^{2}-\frac{\varepsilon}{2}) 
+ \frac{g^{2}a}{2}\varepsilon\sum_{j=1}^{N-1} \sum_{l=1}^{j} \hat{\sigma} _{l}^{z} \notag \\
&&+ \frac{g^{2}a}{4}\sum_{j=1}^{N-1} \sum_{l=1}^{j} \sum_{k=1}^{j}(-1)^{k}\hat{\sigma} _{l}^{z} 
+  \frac{g^{2}a}{8}\sum_{j=1}^{N-1} \sum_{l=1}^{j} \sum_{k=1}^{j}\hat{\sigma} _{l}^{z}\hat{\sigma} _{k}^{z}
\notag \\
\end{eqnarray}
As shown in the Eq.(\ref{H_pauli}), the Schwinger model Hamiltonian can be written as the sum of local Hamiltonians with the form of tensor product of Pauli operators of which the expectation can be measured each one alone. The number of local Hamiltonians is proportional to $N^3$. Therefore,the expectation of the Hamiltonian can  be obtained in polynomial time. 

\section{}
As shown in FIG.2 , we prepare $\left | \psi_0   \right \rangle =\left | 0_{1}  \right \rangle \left |  0_{2} \right \rangle$ as initial state and apply a $Rx(\theta_i)$ gate to the first qubit . Because $Rx(\theta_i)=e^{i\theta_i \sigma_i^x }=\cos \theta_iI+i \sin \theta_i\sigma_i^x$ , $Rx(\theta_i)\left | 0_{1}  \right \rangle=\cos\theta_i\left | 0_{1}  \right \rangle+i \sin\theta_i\left | 1_{1}  \right \rangle$. The state of the subsystem becomes
\begin{eqnarray}
\left | \psi_1   \right \rangle=\cos\theta_i\left | 0_{1}  \right \rangle \left | 0_{2}  \right \rangle+i \sin\theta_i\left | 1_{1}  \right \rangle   \left | 0_{2}  \right \rangle.
\end{eqnarray}
Then we apply a CNOT gate to the two qubits, the state of the subsystem becomes
\begin{eqnarray}
\left | \psi_2   \right \rangle&&=CNOT\left | \psi_1   \right \rangle \notag \\
&&=\cos\theta_i\left | 0_{1}  \right \rangle \left | 0_{2}  \right \rangle+i \sin\theta_i\left | 1_{1}  \right \rangle   \left | 1_{2}  \right \rangle. 
\end{eqnarray}
So the density operator of the subsystem is 
\begin{eqnarray}
\rho&&=\left | \psi_2   \right \rangle \left \langle \psi_2 \right| \notag \\
&&=(\cos\theta_i\left | 0_{1}  \right \rangle \left | 0_{2}  \right \rangle+i \sin\theta_i\left | 1_{1}  \right \rangle   \left | 1_{2}  \right \rangle) \notag \\
&&(\cos\theta_i\left \langle 0_{1}  \right | \left \langle 0_{2}  \right |-i \sin\theta_i\left \langle 1_{1}  \right | \left \langle 1_{2}  \right |) 
\end{eqnarray}
Finally , trace out the first qubit, we get the reduced density operator of the subsystem
\begin{eqnarray}
\label{mix_state_single}
&&\rho_i(\theta_i)=tr_1(\rho) \notag \\
&&=\left \langle 0_{1}  \right |(\cos\theta_i\left | 0_{1}  \right \rangle \left | 0_{2}  \right \rangle+i \sin\theta_i\left | 1_{1}  \right \rangle   \left | 1_{2}  \right \rangle) \notag \\
&&(\cos\theta_i\left \langle 0_{1}  \right | \left \langle 0_{2}  \right |-i \sin\theta_i\left \langle 1_{1}  \right | \left \langle 1_{2}  \right |) \left | 0_{1}  \right \rangle \notag \\
&&+\left \langle 1_{1}  \right |(\cos\theta_i\left | 0_{1}  \right \rangle \left | 0_{2}  \right \rangle+i \sin\theta_i\left | 1_{1}  \right \rangle   \left | 1_{2}  \right \rangle) \notag \\
&&(\cos\theta_i\left \langle 0_{1}  \right | \left \langle 0_{2}  \right |-i \sin\theta_i\left \langle 1_{1}  \right | \left \langle 1_{2}  \right |) \left | 1_{1}  \right \rangle \notag \\
&&=\cos^2\theta_i\left | 0_{2}  \right \rangle \left \langle 0_{2} \right|+\sin^2\theta_i\left | 1_{2}  \right \rangle \left \langle 1_{2} \right|
\end{eqnarray}
According to Eq.(\ref{mix_state_single}), with the circuit in Fig.(\ref{Fig:one_qubit_mix}), we get a mixed state 
\begin{eqnarray}
\rho_i(\theta_i)=\cos^2\theta_i\left | 0  \right \rangle \left \langle 0 \right|+\sin^2\theta_i\left | 1  \right \rangle \left \langle 1 \right|
\end{eqnarray}

\bibliographystyle{apsrev4-2}
\bibliography{tensionT}

\end{document}